\documentclass[a4paper]{panl}
\usepackage{cite}
\usepackage{wrapfig}
\usepackage{graphicx}
\usepackage{amssymb}
\usepackage{amsfonts}
\usepackage{amsmath}
\usepackage{longtable}
\usepackage{rotating}
\usepackage{lscape}
\usepackage{epsfig}
\usepackage{multirow}

\usepackage{lineno}

\originalTeX
\begin{document}

\title{On the new possibility of pulse accumulation of UCN in a trap}
\maketitle
\authors{A.I.\,Frank$^{a,}$\footnote{E-mail: frank@jinr.ru},
G.V.\,Kulin$^{a,}$\footnote{E-mail: kulin@jinr.ru},
M.A.\,Zakharov$^{a,}$}
\setcounter{footnote}{0}
\from{$^{a}$\,Joint Institute for Nuclear Research, Dubna, Russia}


\begin{abstract}
The paper considers the concept of an ultracold neutron source (UCN) based on the deceleration of very cold neutrons (VCN) by a local decelerating device. As the latter, it is proposed to use a gradient spin flipper. It is shown that in this case, the flux of VCNs, which after deceleration are converted into the UCN, has a pulse structure. In this case, the duration of neutron bunches can be significantly less than their repetition period. Accordingly, the density of the neutron flux in the bunch will significantly exceed the average value. This opens up the possibility of pulse filling of the UCN trap, without preliminary time focusing.
\end{abstract}
\vspace*{6pt}
\noindent

PACS: 29.25.Dz Neutron sources; 03.75.Be Atom and neutron optics
\label{sec:intro}
\section*{INTRODUCTION}

It is known that ultracold neutrons (UCN) were first observed by Shapiro's group in an experiment performed at a reactor with an average power of 6 kW~\cite{Luschikov1969,Strelkov2015}. Probably, it was then that there was an understanding of the importance of the fact that the pulse density of the UCN generated by a periodic source can significantly exceed the average value. The question arose how to take advantage of this circumstance. A possible solution to this problem was soon proposed in~\cite{Shapiro1971}. It consists in filling the UCN trap only during the pulse and effectively isolating it the rest of the time. Ideally, when there are no losses, the density of the UCN in the trap will correspond to the pulsed neutron density, which may be several orders of magnitude higher than the average in time.

Unfortunately, this idea has not yet been implemented, although the problem of using pulsed rather than medium UCN density has become even more urgent due to the creation of new pulsed neutron sources~\cite{Anghel2009, Saunders2013, Karch2014}. The IBR 2 – IBR 2M pulse reactor \cite{Ananiev1977, Aksenov2009} with an average power of 1.5–2 MW and a pulse flux of about $10^{16}$ n/cm$^2$ has been successfully operating in Dubna for many decades. The design of a new "Neptune" reactor \cite{Lopatkin2020, Aksenov2021} with a significantly large pulse flux is underway. The construction of the European pulsed neutron source (ESS) is also close to completion \cite{Garoby2018}.

The implementation of the idea of pulse filling of the trap is hindered by the fact that in practice it is remote from the moderator due to the presence of biological protection. In this case,  it is necessitates the appearance of a transport neutron guide several meters long, feeding the trap. The placement of an insulating valve near the moderator -- the source of the UCN, causes the neutron guide to become part of this trap. Due to the small transverse size of the neutron guide, the frequency of neutron collisions against its walls is large enough, which greatly reduces the storage time of the UCN in the trap–neutron guide system and significantly reduces the density of neutrons accumulated in the trap. Placing the valve at the entrance to the trap, several meters away from the source, is useful only in the case of sources with a low repetition rate~\cite{Anghel2009, Saunders2013, Karch2014}. For sources with a repetition rate of several hertz, the spread of the UCN transit times from the source to the trap will exceed the intervals between pulses, and the presence of a valve at the entrance to the trap does not make sense.

To solve the problem of pulsed filling of a remote trap, it was proposed to use a special device - a time lens that dose-changes the energy of neutrons as they come to the lens~\cite{Frank1996, Frank2000}. Such a device makes it possible to restore the pulsed structure of the neutron beam immediately before entering the trap. An important question is the method of changing the neutron energy according to a given time law. In this regard, in ~\cite{Frank1996, Frank2000} it was proposed to turn to quantum nonstationary phenomena. Among the latter, the phase modulation of a neutron wave, across the direction of propagation of which a phase diffraction grating moves, and the resonant the neutron spin flip in a magnetic field were considered.

Later, nonstationary diffraction of the UCN by a moving grating was observed in the experiment~\cite{FrankPLA2003} and some time later, in experiments with a moving grating, the effect of focusing in time was also demonstrated~\cite{FrankJETPL2003, Balashov2004}. The possibility of time focusing based on the resonant the neutron spin flip has also found its experimental confirmation~\cite{Arimoto2012, Imajo2021}.

The concept of a UCN source on a periodic pulsed reactor, based on the use of the time lens with pulse filling of the UCN trap, was considered in a recent paper~\cite{Frank2022}. A similar approach was proposed in~\cite{Nesvizhevsky2022}, in which it was proposed to focus very cold neutrons (VCN) with velocities of about 50 m/s, followed by their deceleration in a escaping trap. Such a deceleration method was proposed in~\cite{Summhammer1986}, but has not yet been applied in practice.

The extraction of neutrons with higher speeds than those of the UCN from the moderator-converter provides better conditions for the transportation of neutrons and allows to use  more efficient converter. In the UCN source of the Institut Laue-Langevin~\cite{Steyerl1986}, neutrons are slowed down rising to a height of several meters, followed by Doppler "cooling" when reflected from an escaping mirror. The deceleration of neutrons in the Earth gravity field during their transportation in a vertical neutron guide was also successfully used in the UCN sources at the WWR-M reactor of PNPI~\cite{Altarev1986}. However, in the case of pulsed neutron generation, the deceleration of VCN may lead to some new and important consequences. The present work is devoted to their discussion.

\label{sec:idea}
\section*{DECELERATION OF NEUTRONS GENERATED BY A PULSED SOURSE USING A LOCAL DEVICE}

We will consider a UCN source in which, for a relatively short time, pulse generation of very cold neutrons occurs and their subsequent transportation through a mirror neutron guide.

Suppose that a device slowing down neutrons is used, the purpose of which is to produce ultracold neutrons, whose energy after deceleration is small enough so that they can be stored in a material trap. Let's call this device a decelerator to avoid the term "moderator" widely used in neutron physics. In contrast to the case considered in~\cite{Nesvizhevsky2022}, we will assume that the deceleration of neutrons by the decelerator occurs in a relatively short section of their transport in close proximity to the trap.

To store in the trap the neutrons obtained due to deceleration, their full velocity $V$ should not exceed the boundary velocity of the trap matter $V<V_b=\sqrt{2U/m}$, where $U$ is the effective potential of the trap walls.

Since before getting into the trap, neutrons must pass a considerable path in the neutron guide, which we assume to be a mirror, the transverse velocity of neutrons normal to the surface of the walls of the latter is limited by the boundary energy value of the neutron guide walls $E_{gd}$, so that $v_\perp <\sqrt{2E_{gd}/m}$. Having obtained a limit for the total and transverse velocity of neutrons capable of being stored in a trap, we thereby obtained a limit for the longitudinal velocity of such neutrons directed along the $Z$ axis of the neutron guide.

\begin{equation}
v_{z\text{fin}}=\sqrt{V_b^2-v_\perp^2}.
\label{eq:eq1}
\end{equation}

The decelerator, which forms the neutron flux immediately before neutrons enter the trap, changes the neutron energy by a certain amount $E_D$. If it is constructed correctly, then this change in kinetic energy is mainly due to a change in the longitudinal velocity of neutrons.

Being interested in the future only in the distribution of the longitudinal velocity of neutrons and the kinetic energy associated with it, we will skip below the $z$ index of the quantities we are interested in. Then the energy of neutrons entering the trap and being able to be stored in it lies in the range from zero to $E_\text{fin}=m v_\text{fin}^2/2$. Before deceleration , the energy of these neutrons should be in the range $E_D<E<E_\text{fin}+E_D$. At a sufficiently large energy value $E_D$, the range of neutron energies that can be trapped after deceleration can be much smaller than the energy itself $\delta E\approx E_\text{fin}\ll E_D$. But this means that the spread of the flight times $\delta t$ from the pulsed source to the decelerator and, accordingly, to the trap, of the "useful" neutrons of interest to us are not only small, but may be much shorter than the time of flight itself $t=L/V$, where $L$ is the length of the transport neutron guide
\begin{equation}
\frac{\delta t}t=\frac{\delta V}V\simeq\frac{\delta E}{2E}\simeq\frac{E_\text{fin}}{2E_D}\ll 1.
\label{eq:eq2}
\end{equation}

Under good conditions, $\delta t$ can also be significantly less than the pulse repetition period of the reactor $T$. In this case, the flux of "useful" neutrons, which after deceleration will be converted to the UCN, will have a pulsed structure, although during transport through the neutron guide, the pulse duration will inevitably increase due to velocity dispersion $\delta V$.

Let us now consider the question of the fluxes of UCN coming into the trap. Obviously, the process of deceleration does not affect the number of neutrons in any way and, consequently, the decelerator itself does not change the flux of neutrons with the corresponding spin projection. Therefore, forgetting about polarization for now, and assuming that the transmission of neutron guides for neutrons with the above velocity distribution is ideal, let's compare the flux of UCN $\Phi_{1z}$ that would enter the trap directly from the source with the flux of "useful" neutrons  $\Phi_{2z}$ coming from the source to the decelerator and only then into the trap. Assuming that the distribution over the "normal" velocities $v_\perp$is the same in both cases, we will continue to omit the $z$ index, being interested only in the components of the velocities directed along the beam and the corresponding fluxes.

The flux $\Phi_{1}$ is carried by neutrons with velocities ranging from zero to $v_\text{fin}$, and the flux $\Phi_{2}$ is carried by neutrons with velocities from $V_1=\sqrt{2E_D/m}$ to $V_2=\sqrt{2\left(E_D+E_\text{fin}\right)/m}$. Since the UCN energy is very small, then in the Maxwell distribution for the velocity distribution of the flux in the source the linear approximation is valid $d\Phi(V)=nVdV$, where $n$ is the neutron density. Then for both fluxes we have

\begin{equation}
\Phi_{1}=n\int^{\sqrt{2E_\text{fin}/m}}_0 VdV,\;\;\;\Phi_{2}=n\int^{\sqrt{2\left(E_D+E_\text{fin}\right)/m}}_{\sqrt{2E_D/m}} VdV.
\label{eq:eq3}
\end{equation}

It is easy to see that $\Phi_{1}=\Phi_{2}$.

Thus, neutrons entering the trap directly from the source and neutrons obtained by converting from VCN to UCN carry the same flux, but have a significantly different temporal and spatial structure. In the first case, the spread of the flight times $\delta t$ is much larger than the pulse repetition period $T$. In this case, the pulse structure practically disappears and an essentially uniform flux enters the trap, which corresponds to the average value of the neutron density. In the second case, when the duration of the bunch is significantly less than the pulse repetition period $\delta t/T\ll 1$, the length of the bunches is less than the distance between them. Accordingly, the neutron density in the bunch exceeds the average by a value of $G=T/\delta t$.

\label{sec:estimates}
\section*{ADIABATIC SPIN-FLIPPER FOR NEUTRON DECELERATION. QUANTITATIVE ESTIMATES}

Let us make some estimates now. For certainty, as a decelerator, we consider a flipper in which a spin flip occurs under the action of an alternating high-frequency field directed perpendicular to a large permanent field. As such, the so-called adiabatic or gradient flipper can be used~\cite{Egorov1974, Luschikov1984, Grigoriev1997}. Passing the flipper, the neutron energy changes by an amount $E_D=2\mu B$, where $\mu$ is the magnetic moment of the neutron and $B$ is the magnitude of the permanent magnetic field, and under good conditions this energy change is mainly due to a change in the longitudinal velocity of the neutrons.

The energy change at the neutron spin reversal in the such flipper was demonstrated in~\cite{Weinfurter1988}, and the possibility of creating a flipper with the permanent field of the order of 1~T was demonstrated in~\cite{Arimoto2012, Imajo2021}. There is no reason to doubt that it is possible to create a flipper with a field of the order of 15-20~T, which is achievable in modern superconducting systems.

For the final velocity $v_\text{fin}$, we assume a value of 3 m/s, so that the total velocity in a circular neutron guide with a boundary energy of 5.7~m/s does not exceed 6.5~m/s. This velocity corresponds to an energy of the order of 50~neV. For the magnitude of the magnetic field $B$, in which the spin flip occurs, we take the value of 20~T.

When the spin is reversed in such a field, the energy changes by the value $E_D = 2.4\times10^3$~neV. Therefore, the spread of neutron energies, whose longitudinal velocity after deceleration will not exceed 3~m/s, should be about 50~neV, while the energy itself will be somewhat greater than $E_D$. This energy corresponds to the neutron velocity of 21~m/s.

From formula ~\eqref{eq:eq2} follows the estimate

\begin{equation}
\frac{\delta t}t\approx 0.01.
\label{eq:eq4}
\end{equation}

Assuming for a rough estimate the length of the neutron guide is $L\approx10$~m, we obtain for the time of flight and its dispersion the values $t=0.48$~s and $\delta t\approx 5$~ms.

The last of these values determines the duration of the bunch of "useful" neutrons at the entrance to the flipper-decelerator. This duration $\delta t$ may be significantly less than the pulse repetition period of the source $T$. In particular, for the IBR-2 reactor, $T = 200$~ms.

At the same time, at the direct transport of neutrons from the source to the trap, the spread of the flight times of the order of the flight time itself is $\delta t \approx L/v_\text{fin} \approx3$~s, which is an order of magnitude greater than the repetition period $T$ and the density of neutrons reaching the trap in this case corresponds to the average value. Remind that if the fluxes are equal, that is following from~\eqref{eq:eq3}l, the compression of a bunch of neutrons by their local deceleration means an increase in the neutron density proportional to the value of $G$.

Above, we assumed that the deceleration time in the flipper is the same for all neutrons. This is obviously not the case. The process of neutron deceleration in the flipper is caused by the slowing down of neutrons in the increasing magnetic field when entering it, and then, after the spin flip, the same slowing down in the decreasing field when exiting the flipper. These two stages are complemented by some time of flight in the weakly inhomogeneous field of the flipper itself. The deceleration time and its dispersion are determined by the design of the flipper and the range of initial and final neutron velocities. This should be taken into account at its designing. Rough estimates based on the assumption of the constant neutron deceleration in a region $l$ of an inhomogeneous magnetic field lead to the magnitude of the dispersion of the deceleration times is $\delta t\approx 2 l V_2/V_1^2\approx 10\div 15$~ms, where $V_1$ and $V_2$ are the neutron velocities before and after deceleration. This value can be reduced if neutrons with the lowest velocities are excluded from consideration, which will lead to a very insignificant loss in neutron flux.

\label{sec:conclusion}
\section*{CONCLUSION}

Thus, it is shown that decelerating the neutrons generated by a pulsed source, with the help of a local device, it is possible to obtain a noticeable gain in the pulse density of the UCN and without time focusing. The question of the combination of these two approaches, partially raised in~\cite{Nesvizhevsky2022}, probably requires additional analysis.

The authors are grateful to E.~V. Lychagin, O.~V. Karamyshev, S.~V. Mironov, A.~Yu. Muzychka and M.~S. Novikov for useful discussions.



\begin{thebibliography}{10}
\def\selectlanguageifdefined#1{
\expandafter\ifx\csname date#1\endcsname\relax
\else\selectlanguage{#1}\fi}
\providecommand*{\href}[2]{{\small #2}}
\providecommand*{\url}[1]{{\small #1}}
\providecommand*{\BibUrl}[1]{\url{#1}}
\providecommand{\BibAnnote}[1]{}
\providecommand*{\BibEmph}[1]{\emph{#1}}
\ProvideTextCommandDefault{\cyrdash}{\hbox to.8em{--\hss--}}
\providecommand*{\BibDash}{\ifdim\lastskip>0pt\unskip\nobreak\hskip.2em\fi
\cyrdash\hskip.2em\ignorespaces}

\bibitem{Luschikov1969}
\selectlanguageifdefined{english}
\BibEmph{Luschikov V.I., Pokotilovsky {Yu}.N., Strelkov A.V., Shapiro F.L.}
  {Observation of ultracold neutrons}~// Journal of Experimental and
  Theoretical Physics Letters. \BibDash
\newblock 1969. \BibDash
\newblock V.~9, no.~1. \BibDash
\newblock P.~23--26.

\bibitem{Strelkov2015}
\selectlanguageifdefined{english}
\BibEmph{Strelkov A.V.} {The history of the discovery of ultracold neutrons}~//
  {Shapiro F. L. Collection of works. Neutron Studies}. \BibDash
\newblock Moscow: Nauka, 2015. \BibDash
\newblock P.~362--364. \BibDash
\newblock {in Russian}.

\bibitem{Shapiro1971}
\selectlanguageifdefined{english}
\BibEmph{Shapiro F.L.} {Remarks on the measurement of phases of structural
  amplitudes in neutron diffraction and on the accumulation of neutrons}~//
  Physics of Elementary Particles and Atomic Nuclei. \BibDash
\newblock 1971. \BibDash
\newblock V.~2, no.~4. \BibDash
\newblock P.~975--979. \BibDash
\newblock {in Russian}.

\bibitem{Anghel2009}
\selectlanguageifdefined{english}
\BibEmph{Anghel A., Atchison F., Blau B., . et al.} {The PSI ultra-cold
  neutron source}~// Nuclear Instruments \& Methods in Physics Research Section
  A-accelerators Spectrometers Detectors and Associated Equipment. \BibDash
\newblock 2009. \BibDash
\newblock V. 611. \BibDash
\newblock P.~272--275.

\bibitem{Saunders2013}
\selectlanguageifdefined{english}
\BibEmph{Saunders A., Makela M., Bagdasarova Y. et al.} {Performance of
  the Los Alamos National Laboratory spallation-driven solid-deuterium
  ultra-cold neutron source}~// The Review of scientific instruments. \BibDash
\newblock 2013. \BibDash
\newblock V.~84. \BibDash
\newblock P.~013304.

\bibitem{Karch2014}
\selectlanguageifdefined{english}
\BibEmph{Karch J., Sobolev {Yu}., Beck M. et al.} {Performance of the
  solid deuterium ultra-cold neutron source at the pulsed reactor TRIGA
  Mainz}~// The European Physical Journal A. \BibDash
\newblock 2014. \BibDash
\newblock V.~50, no.~78.

\bibitem{Ananiev1977}
\selectlanguageifdefined{english}
\BibEmph{Ananiev V.D., Blokhintsev D.I., Bulkin {Yu}.M. et al.} {IBR}-2 -
  pulsed reactor for neutron investigations~// Pribory i Tekhnika
  Ehksperimenta. \BibDash
\newblock 1977. \BibDash
\newblock V.~5. \BibDash
\newblock P.~17--35.

\bibitem{Aksenov2009}
\selectlanguageifdefined{english}
\BibEmph{Aksenov V.L.} Pulsed nuclear reactors in neutron physics~// Physics
  Uspekhi. \BibDash
\newblock 2009. \BibDash
\newblock V.~52. \BibDash
\newblock P.~406--413.

\bibitem{Lopatkin2020}
\selectlanguageifdefined{english}
\BibEmph{Lopatkin A.V., Tret'yakov I.T., Romanova N.V. et al.} Concept
  of a {N}ew {H}igh-{F}lux {P}eriodic-{P}ulse {S}ource of {N}eutrons {B}ased on
  {N}eptunium~// Atomic Energy. \BibDash
\newblock 2021. \BibDash
\newblock V. 129, no.~4. \BibDash
\newblock P.~227--230.

\bibitem{Aksenov2021}
\selectlanguageifdefined{english}
\BibEmph{Aksenov V.L., Rzyanin M.V., Shabalin E.P. et al.} Research
  {R}eactors at {JINR}: {L}ooking into the {F}uture~// Physics of Particles and
  Nuclei. \BibDash
\newblock 2021. \BibDash
\newblock V.~52, no.~6. \BibDash
\newblock P.~1019--1032.

\bibitem{Garoby2018}
\selectlanguageifdefined{english}
\BibEmph{Garoby R., Vergara A., Danared H. et al.} Corrigendum: {T}he
  {E}uropean {S}pallation {S}ource {D}esign (2018 {P}hys. {S}cr. 93 014001)~//
  Physica Scripta. \BibDash
\newblock 2018. \BibDash nov. \BibDash
\newblock V.~93, no.~12. \BibDash
\newblock P.~129501.

\bibitem{Frank1996}
\selectlanguageifdefined{english}
\BibEmph{Frank A.I., Gahler R.} Neutron {T}ime {F}ocusing~// Proceedings of
  {IV} {I}nternational {S}eminar on {I}nteraction of {N}eutrons with {N}uclei~/
  JINR. \BibDash
\newblock E3-96-336. \BibDash
\newblock Dubna~: JINR, 1996. \BibDash
\newblock P.~308--.

\bibitem{Frank2000}
\selectlanguageifdefined{english}
\BibEmph{Frank A.I., Gahler R.} Time focusing of neutrons~// Physics of Atomic
  Nuclei. \BibDash
\newblock 2000. \BibDash
\newblock V.~63. \BibDash
\newblock P.~545--547.

\bibitem{FrankPLA2003}
\selectlanguageifdefined{english}
\BibEmph{Frank A.I., Balashov S.N., Bondarenko I.V. et al.} Phase modulation
  of a neutron wave and diffraction of ultracold neutrons on a moving
  grating~// Physics Letters A. \BibDash
\newblock 2003. \BibDash
\newblock V. 311, no.~1. \BibDash
\newblock P.~6--12.

\bibitem{FrankJETPL2003}
\selectlanguageifdefined{english}
\BibEmph{Frank A.I., Geltenbort P., Kulin G.V., Strepetov A.N.} A quantum time
  lens for ultracold neutrons~// Journal of Experimental and Theoretical
  Physics Letters. \BibDash
\newblock 2003. \BibDash
\newblock V.~78. \BibDash
\newblock P.~188--192.

\bibitem{Balashov2004}
\selectlanguageifdefined{english}
\BibEmph{Balashov S.N., Bondarenko I.V., Frank A.I. et al.} Diffraction
  of ultracold neutrons on a moving grating and neutron focusing in time~//
  \href{http://dx.doi.org/10.1016/j.physb.2004.04.038}{Physica B: Condensed
  Matter}. \BibDash
\newblock 2004. \BibDash
\newblock V. 350, no.~1. \BibDash
\newblock P.~246--249. \BibDash
\newblock Proceedings of the Third European Conference on Neutron Scattering
  URL:
  \BibUrl{https://www.sciencedirect.com/science/article/pii/S0921452604006118}.

\bibitem{Arimoto2012}
\selectlanguageifdefined{english}
\BibEmph{Arimoto Y., Gertenbort P., Imajo S. et al.} Demonstration of
  focusing by a neutron accelerator~// Physical Review A. \BibDash
\newblock 2012. \BibDash Aug. \BibDash
\newblock V.~86. \BibDash
\newblock P.~023843.

\bibitem{Imajo2021}
\selectlanguageifdefined{english}
\BibEmph{Imajo S., Iwashita Y., Mishima K. et al.} Ultracold {N}eutron
  {T}ime {F}ocusing {E}xperiment and {P}erformance {E}valuation of an
  {I}mproved {UCN} {R}ebuncher at {J}-{PARC}/{MLF}~// JPS Conference
  Proceedings. \BibDash
\newblock 2021. \BibDash
\newblock V.~33. \BibDash
\newblock P.~0111091.

\bibitem{Frank2022}
\selectlanguageifdefined{english}
\BibEmph{Frank A.I., Kulin G.V., Rebrova N.V., Zakharov M.A.} On the
  {P}ossibility of {C}reating a {UCN} {S}ource at a {P}eriodic {P}ulsed
  {R}eactor~// Physics of Particles and Nuclei. \BibDash
\newblock 2022. \BibDash
\newblock V.~53, no.~1. \BibDash
\newblock P.~33--44.

\bibitem{Nesvizhevsky2022}
\selectlanguageifdefined{english}
\BibEmph{Nesvizhevsky N.N., Sidorin A.O.} Production of {U}ltracold {N}eutrons
  in an {E}scaping {D}ecelerating {T}rap~// Physics of Particles and Nuclei
  Letters. \BibDash
\newblock 2022. \BibDash
\newblock V.~19. \BibDash
\newblock P.~162--175.

\bibitem{Summhammer1986}
\selectlanguageifdefined{english}
\BibEmph{Summhammer J., Niel L., Rauch H.} Focusing of {P}ulsed {N}eutrons by
  {T}raveling {M}agnetic {P}otentials~// Zeitschrift f\"ur Physik B. \BibDash
\newblock 1986. \BibDash
\newblock V.~62. \BibDash
\newblock P.~269--278.

\bibitem{Steyerl1986}
\selectlanguageifdefined{english}
\BibEmph{Steyerl A., Nagel H., Schreiber F.-X. et al.} A {N}ew {S}ource
  of {C}old and {U}ltracold {N}eutrons~// Physics Letters A. \BibDash
\newblock 1986. \BibDash
\newblock V. 116. \BibDash
\newblock P.~347--352.

\bibitem{Altarev1986}
\selectlanguageifdefined{english}
\BibEmph{Altarev I.S., Borisov N.V., Borovikova A.B. et al.} Search for
  an electric dipole moment of the neutron~// JETP Letters. \BibDash
\newblock 1986. \BibDash
\newblock V.~44, no.~8. \BibDash
\newblock P.~460--465.

\bibitem{Egorov1974}
\selectlanguageifdefined{english}
\BibEmph{Egorov A.I., Lobashov V.M., Nazarenko V.A. et al.} Production,
  storage, and polarization of ultracold neutrons~// Soviet Journal of Nuclear
  Physics. \BibDash
\newblock 1974. \BibDash
\newblock V.~19, no.~2. \BibDash
\newblock P.~147--152.

\bibitem{Luschikov1984}
\selectlanguageifdefined{english}
\BibEmph{Luschikov V.I., Taran Y.V.} On the calculation of the neutron
  adiabatic spin-flipper~// Nuclear Instruments and Methods in Physics
  Research. \BibDash
\newblock 1984. \BibDash
\newblock V. 228. \BibDash
\newblock P.~159--160.

\bibitem{Grigoriev1997}
\selectlanguageifdefined{english}
\BibEmph{Grigoriev S.V., Okorokov A., Runov V.} Peculiarities of the
  construction and application of a broadband adiabatic flipper of cold
  neutrons~// Nuclear Instruments and Methods in Physics Research A. \BibDash
\newblock 1997. \BibDash
\newblock V. 384. \BibDash
\newblock P.~451--456.

\bibitem{Weinfurter1988}
\selectlanguageifdefined{english}
\BibEmph{Weinfurter H., Badurek G., Rauch H., Schwahn D.} Inelastic action of a
  gradient radio-frequency neutron spin flipper~//
  \href{http://dx.doi.org/10.1007/BF01312135}{Zeitschrift f\"ur Physik B
  Condensed Matter}. \BibDash
\newblock 1988. \BibDash
\newblock V.~72, no.~2. \BibDash
\newblock P.~195--201. \BibDash
\newblock URL: \BibUrl{https://doi.org/10.1007/BF01312135}.

\end{thebibliography}
\end{document}